%% file: conference_041818.tex
\documentclass[sigconf]{acmart}

\usepackage{cite}
\usepackage{adjustbox}
\usepackage{graphicx,lineno}
\usepackage{algorithm}
\usepackage{algorithmic}
\usepackage{multirow}
\usepackage[tight,footnotesize]{subfigure}
\usepackage{footmisc}
\usepackage{footnpag}
\usepackage{paralist}
\usepackage{color}
\usepackage{perpage}
\usepackage{verbatim}
\usepackage{enumitem}
\usepackage{listings}
\setcopyright{none}
\acmConference[Conference'21]{conference}{March, 2021}{}
\acmYear{2021}

\begin{document}


\title{SDC Resilient Error-bounded Lossy Compressor
}

\author{Sihuan Li}
\affiliation{%
  \institution{University of California, Riverside}
  \streetaddress{900 University Avenue}
  \city{Riverside}
  \state{CA}
  \postcode{92521}
}
\email{sli049@ucr.edu}

\author{Sheng Di}
\affiliation{%
  \institution{Argonne National Laboratory}
  \streetaddress{9700 Cass Avenue}
  \city{Lemont}
  \state{IL}
  \postcode{60439}
}
\email{sdi1@anl.gov}

\author{Kai Zhao}
\affiliation{%
  \institution{University of California, Riverside}
  \streetaddress{900 University Avenue}
  \city{Riverside}
  \state{CA}
  \postcode{92521}
}
\email{kzhao016@ucr.edu}

\author{Xin Liang}
\affiliation{%
  \institution{Oak Ridge National Laboratory}
  \streetaddress{1 Bethel Valley Rd}
  \city{Oak Ridge}
  \state{TN}
  \postcode{37830}
}
\email{liangx@ornl.gov}


\author{Zizhong Chen}
\affiliation{%
  \institution{University of California, Riverside}
  \streetaddress{900 University Avenue}
  \city{Riverside}
  \state{CA}
  \postcode{92521}
}
\email{chen@cs.ucr.edu}

\author{Franck Cappello}
\affiliation{%
  \institution{Argonne National Laboratory}
  \streetaddress{9700 Cass Avenue}
  \city{Lemont}
  \state{IL}
  \postcode{60439}
}
\email{cappello@mcs.anl.gov}
\input{tex/abstract.tex}

\maketitle


\input{tex/introduction.tex}
\input{tex/relatedwork.tex}
\input{tex/problemform.tex}
\input{tex/sz_sdc_analysis.tex}

\input{tex/methodology.tex}
\input{tex/evaluation.tex}
\input{tex/conclusion.tex}

\bibliographystyle{ACM-Reference-Format}
\bibliography{tex/ref.bib}

\end{document}

%% file: tex/abstract.tex
\begin{abstract}

Lossy compression is one of the most important strategies to resolve the big science data issue, however, little work was done to make it resilient against silent data corruptions (SDC). In fact, SDC is becoming nonnegligible because of exa-scale computing demand on complex scientific simulations with vast volume of data being produced or in some particular instruments/devices (such as interplanetary space probe) that need to transfer large amount of data in an error-prone environment. In this paper, we propose an SDC resilient error-bounded lossy compressor upon the SZ compression framework. Specifically, we adopt a new independent-block-wise model that decomposes the entire dataset into many independent sub-blocks to compress. Then, we design and implement a series of error detection/correction strategies based on SZ. We are the first to extend algorithm-based fault tolerance (ABFT) to lossy compression. Our proposed solution incurs negligible execution overhead without soft errors. It keeps the correctness of decompressed data still bounded within user's requirement with a very limited degradation of compression ratios upon soft errors.
\end{abstract}

%% file: tex/introduction.tex
\section{Introduction}

Error-bounded lossy compressors \citep{zfp,fpzip,sz,sz2,sz3} have been effective in significantly reducing large volumes of data produced by scientific simulations \citep{ssem,hacc, impact-hpdc14} or instruments/devices \citep{APSU,LCLSII}, while controlling the data distortion based on user's requirement. Accordingly, error-bounded lossy compression has been thought of as one of the best ways to resolve today's big science data issue.  

Silent data corruptions (SDC), however, are nonnegligible when running lossy compressors, as discussed below. 
\begin{itemize}
    \item If one lossy compressor is employed by a high performance computing (HPC) application, it will likely need to deal with a vast volume of data produced by extreme-scale simulations. Various possible failures/errors have to be taken into account.
    Many existing solutions such as multi-level checkpointing/restart (CR) mechanism focus only on fail-stop issues that are perceived by hardware or operating systems.
    By contrast, the soft errors, a.k.a. silent data corruption (SDC), may change the data in memory, cache or even register silently, because of inevitably unexpected malfunctions in the system components. Such errors are more dangerous than fail-stop issues, because they may cause biased results in the end of simulation silently.
    \item Remote sensor technology continues to increase in fidelity for space systems, so large amounts of data are being collected by orbiting satellites or space vehicles and transmitted to other stations (e.g., ground stations, other satellites). However, the devices (such as interplanetary space probe) deployed in space would be more error-prone than the regular devices on the earth. To address this issue, some fault tolerance techniques \citep{Lin-ft-space,Jacobs-ft-space} have been proposed for specific algorithms such as matrix multiplication and FFT.
    However, when lossy compressors are used by the space systems to compress image data, the whole compression procedure has to be protected against soft errors. Otherwise, the corrupted data may let scientists miss important findings or draw a misleading conclusion.
\end{itemize}

There are no lossy compressors designed particularly in the consideration of the possible SDCs. Mat Noor and Vladimirova \citep{lossless-ft-space} made a parallel fault-tolerant Integer KLT implementation for lossless hyperspectral image compression on board satellites. Unlike the lossless compression, designing SDC detection/correction method for lossy compression is more challenging since decompressed data will deviate from original data even though there is no SDC. 


In this paper, we propose the SDC resilient lossy compression based on SZ \citep{sz3} - one of the best generic error-bounded lossy compressors for large-scale scientific datasets verified by many studies \citep{sz3, understand-compression-ipdps18}. 
Not only can our solution detect the SDCs during the compression/decompression but it can also automatically correct the SDCs in some cases.

In general, the SDC errors can be classified into two categories, memory error and computation error, and our solution can protect SZ against both of the two errors. The memory error is introduced by soft errors that corrupt a data value in memory from $a$ to $a'$. The computation error is introduced by soft errors in logic unit which yields wrong computation results such as $1+1 = 3$. 

The main idea of this paper is analyzing each subroutine in the SZ framework elaborately and designing a series of fault tolerance strategies carefully, such that the lossy compressor can be protected against SDCs effectively with little overhead. We summarize the detailed contributions as follows.
\begin{itemize}[leftmargin=*]
\item We comprehensively analyze each subroutine of SZ with respect to possible memory/computation errors. The analysis unveils that some parts of SZ are naturally error resilient, 
while other parts are fragile to SDCs. The SDC errors striking these parts may cause wrong decompressed data. Thus, it is critical to protect those parts by specific fault tolerance strategies. 
\item We propose an efficient SDC resilient lossy compression solution based on the SZ compression framework. e reorganize the SZ compression model by dividing each dataset into small blocks and making the compression work totally independent across blocks. Such a design is able to control the impact of SDCs on the decompressed data. On the other hand, we design a series of SDC resilient strategies based on SZ's principle, which can not only detect SDCs in most of cases but also correct SDCs in some cases.
\item We implement our SDC resilient compressor based on our elaborate design. We evaluate its fault tolerance ability in the presence of SDCs and the corresponding overhead in the fault-free situation, as well as the possible impact to the compression quality. We perform the experiments with real-world simulation data across multiple science domains and image data which were taken by New Horizons space probe \citep{new-horizons} in the space. Experiments show that our designed independent-block based compression model has very limited execution overheads ($\leq$10\% in most cases). On the other hand, the experiments also confirm that our fault tolerance solution yields little overhead ($\leq$7.3\% at 2048 cores) and correct decompression results in the presence of soft errors. When injecting one and two errors, respectively, during the compression at runtime, our solution can significantly improve the resilience for SZ (92\% running cases with correct decompressed data compare to only 71.2\% and 47\% of the original SZ). 
\end{itemize}

We organize our paper as follows. In Section \ref{sec:relate}, we discuss related work. In Section \ref{sec:problemform}, we formulate the research problem in terms of the SZ compression framework. In Section \ref{sec:err-analysis-sz}, we provide an in-depth analysis of the fault tolerance ability of SZ 
In Section \ref{sec:methodology}, we present our fault tolerance methodology. Then we evaluate our methods in Section~\ref{sec:experiment}. Lastly we conclude the paper and discuss the future work. 

%% file: tex/relatedwork.tex
\section{Related Work}
\label{sec:relate}

We discuss the related work in two facets: the fault tolerance ability of existing lossy compressors and the existing solutions designed to protect other applications against SDCs.

So far, there have been many lossy compressors \citep{isabela,fpzip,zfp,sz,sz2,sz3,numarck,ssem} developed to significantly reduce the large volume of data produced by scientific simulations. All the lossy compressors, basically, could be classified into two categories - transform-based compression \citep{zfp,ssem} and prediction-based compression \citep{fpzip,sz,sz2,sz3}. 
None of the transform-based compressors are immune to the SDCs. In fact, if the data in the transformed domain are corrupted because of memory or computation error, multiple data values in the original data domain could be affected. As for the prediction-based model, the SDC issue could be also fatal to the reconstruction of data. 
In SZ, for example, if the data prediction on some data point is corrupted silently during the compression, the predicted value on that data point would be inconsistent during the compression and decompression, leading to uncontrolled decompression errors. 

Much work has been done to fight against the memory error and computation error, respectively. From the perspective of hardware, error correcting code (ECC) has been implemented to detect and correct bit flips in memory. 
ECC can correct single-bit flipped memory errors but cannot detect or correct any computation errors. 
Hardware redundancy adopts redundant hardware to execute the same application with the same input and compare the outputs from the different hardwares. Software redundancy means running the same program on the single hardware multiple times and compare the outputs from different runs. Thus, double modular redundancy (DMR) is needed for error detection with $100\%$ overhead and triple modular redundancy (TMR) is needed for error correction with $200\%$ overhead.

Such high overhead of modular redundancy to handle SDCs has motivated algorithm based fault tolerance (ABFT) \citep{abft-huang1984}, which aims to exploit the special characteristics of an application or algorithm to detect and correct soft errors. Despite the fact that ABFT requires a significant algorithm integration effort, the tiny overhead of ABFT  makes it very attractive. Most of the existing ABFT methods, however, focus on popular arithmetic algorithms such as matrix operations \citep{abft-huang1984}. To the best of our knowledge, no ABFT work has been done for lossy compression algorithms, 
which is a significant gap in the context of scientific data compression.

%% file: tex/problemform.tex
\section{Background and Problem Formulation}
\label{sec:problemform}


\subsection{SZ Lossy Compression Framework}
SZ \citep{sz3} is an error-bounded lossy compressor designed for scientific data. According to the recent studies \citep{sz2,sz3,understand-compression-ipdps18}, it can effectively reduce the data size for many scientific simulations, such as climate simulation, cosmological simulation, quantum simulation, and chemical simulation. 

Basically, SZ includes four critical stages during the compression: (1) data prediction, (2) linear-scaling quantization, (3) variable-length encoding, and (4) lossless compression such as Zstd \citep{zstd}. In the data prediction step, SZ \citep{sz16,sz17,sz3} splits the whole dataset into multiple blocks in the size of 6x6x6 and then perform the compression in each block based on two alternative prediction methods - an improved Lorenzo predictor \citep{lorenzo} or linear regression. The second step - linear-scaling quantization  converts each raw data value (such as floating-point value) to an integer index (or quantization bin) based on the user-set error bound and the difference of the predicted value and original value. The remaining two steps are used to reduce the data size by performing Huffman encoding on the quantization bin index array and adopting lossless compression. This may significantly reduce the data size because the distribution of quantization bin indices are likely fairly non-uniform especially when the data are relatively smooth in space.

\subsection{Algorithm based fault tolerance (ABFT)}
\label{sec:abft}
ABFT achieves SDC detection and correction by leveraging the characteristics of the algorithms. In high level explanations, ABFT detects SDCs by checking if some relationship is respected and correct the errors by another introduced set of computation. Each ABFT technique has to be developed for a particular approach composed by one or more algorithms. 
We give an example to illustrate how ABFT detects/corrects soft errors in general. Given an array $a[]$ at timestamp $t_0$, then at a later timestamp $t_1$, one attempts to detect if there was a memory error that corrupted a value in $a[]$ during the period $[t_0, t_1]$. In order to detect the error, we can leverage a checksum (\textit{sum} = $\sum$ \textit{a[i]}). Specifically, we can calculate the sum of $a[]$ at $t_0$ and $t_1$, respectively. Suppose the two calculated sums are denoted by $sum_{t0}$ and $sum_{t1}$, respectively. If $sum_{t0}$$\neq$$sum_{t1}$, we can conclude there must be an SDC error happening to $a[]$ during the period $[t_0, t_1]$.
In order to locate where the SDC error is in the array $a[]$, we can leverage an extra computation: \emph{isum = $\sum$ i*a[i]}. Specifically, assuming the value at index $j$ is corrupted during the time period [$t_0,t_1$], according to $sum_{t_1} - sum_{t_0} = a[j]' - a[j]$ and $isum_{t_1} - isum_{t_0} = j*(a[j]' - a[j])$, one can derive the SDC location index $j$ = $(isum_{t_1} - isum_{t_0})/(sum_{t_1} - sum_{t_0})$. This example illustrates that it is viable to detect and even correct the single-data-point error just by introducing a few more light-weight computations.

\subsection{Error model and assumptions}
\label{sec:errmodel}
We identify the error model in this subsection. In our study, we focus on both memory error and computation error. As for the memory error model, the errors could randomly happen anywhere in the whole memory at any time during the life time of a process in the form of bit-flips. 
As for the computation errors, their impact could appear in the form of bit-flips on the computation results. Similar to other ABFT research, the flow control error (FCE) is beyond the scope of our work because the general solutions are designed on the compiler/instruction/hardware level \citep{cfc-not}. Moreover, it is too difficult to comprehensively detect the FCEs even for professional FCE detection tools according to recent studies \citep{cfc-not}. 
Without loss of generality, we assume that the occurrence probability of multiple computation errors or memory errors is extremely low for one block of data during one compression, since one block is generally very small (such as 10$\times$10$\times$10 in size). Similar to other ABFTs \citep{xinfft,panruosvd,jieyangipdps}, we assume the checksum itself is error free because of its tiny computation time compared with the compression time. 
\subsection{Formulation of SDC Detection Evaluation in SZ }

As mentioned previously, SZ has four stages in the whole course of compression, and we mainly focus on the single-data-point SDC error (either computation error or memory error) happening at each stage, without loss of generality. In addition, we mainly focus on the dominant data structures (i.e., all the data structures taking linear space of the number of data points $N$) that take the majority of memory footprint in SZ because they are the major objects affected by SDCs if any. 
The rest parts (called \emph{negligible space} in the following text) could be considered error free. Which parts taking negligible space will be discussed later in this paper.

The objective of our work is to detect and correct both computational errors and memory errors in each stage of SZ compression as much as possible. There are three important metrics to evaluate our designed SDC resilient lossy compressor, as listed below. 

\begin{itemize}[wide=0pt]
    \item \textit{SDC detection/correction ability}. What kinds of SDCs could be detected or corrected? What is the accuracy and coverage rate of SDC error detection? 
    \item \textit{Computational Overhead}. It is defined as the ratio of the extra computation time to the total original execution time in an error-free situation.
    \item \textit{Impact to Compression Result}. Whether the SDC resilient lossy compressor can still respect the user-specified error bound for the decompressed data? What is the compression overhead: i.e, how much the compression ratio would be degraded under the SDC resilient compressor compared with the original compressor?
\end{itemize}

All the three evaluation metrics can be used to all lossy compressors, which is the first resilience formulation in the context of lossy compression, to the best of our knowledge. 

%% file: tex/sz_sdc_analysis.tex
\section{Resilience Analysis of SZ 2.1}
\label{sec:err-analysis-sz}

In this section, we analyze the resilience 
(SDC detection/correction ability and impact)
of SZ 2.1 based on its principle. 

\begin{figure*}[ht]
    \centering
    \hspace{3mm}
    \includegraphics[width=18cm]{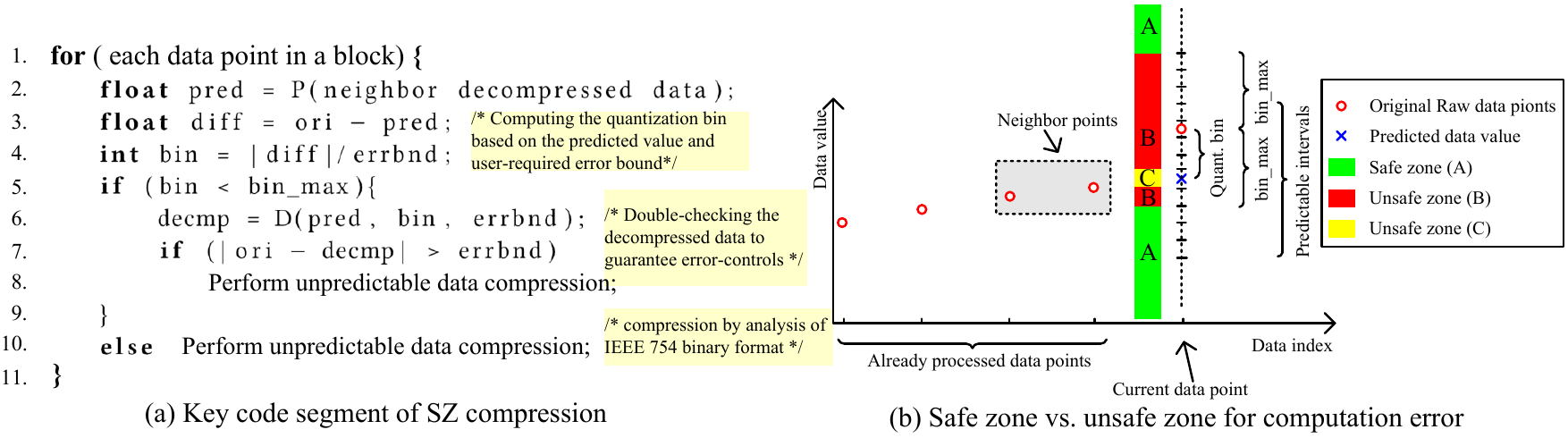}
    \caption{Analysis of fault tolerance ability for SZ with computation error}
    \label{fig:sz-ft-analysis}
\end{figure*}

\vspace{-2mm}
\subsection{SDC Resiliency -- Computation error}
\label{sec: comp err rslc}

We analyze SZ's natural resilience based on when/where the computation error could happen, including calculation of regression coefficients, selecting bestfit predictor by sampling method, and data prediction and calculation of decompressed data, huffman encoding and lossless compression. We call the first two stages  `prediction preparation'.

\subsubsection{SDC resilience in the prediction preparation}
\label{sec:analysis of comp err in reg coeff}
A computation error in prediction preparation stage may only lower compression ratio to a certain extent but it would not affect the correctness of decompressed data (i.e., still strictly respecting error bound).
That is, the decompressed data is still the golden result in spite of the computation error in prediction preparation. In fact, although the computation error may lead to inaccurate regression coefficients or incorrect bestfit predictor selection, exactly the same coefficients/selection will be used for both compression and decompression. The compression ratio could be affected because the data prediction may be less accurate due to the inaccurate coefficients or incorrect predictor selection.

\subsubsection{SDC Resilience in the data prediction and calculation of decompressed data}
\label{sec:resi-data-pred}

Data prediction is the most critical step in SZ. In order to guarantee the error bound, the neighboring data values used to predict each data point during the compression have to be exactly the same values to be used during the decompression. That is, SZ needs to obtain the decompressed data values during compression. We demonstrate the key compression procedure in Figure \ref{fig:sz-ft-analysis} (a), which is conducted in a loop of scanning all data blocks. It involves 5 key steps. 

\begin{enumerate}[leftmargin=*]
    \item Calculate predicted value (line 2).
    \item Compute the difference between the real value and the predicted value (line 3).
    \item Calculate error quantization bins (line 4).
    \item Calculate the decompressed data (line 6) which will be used to predict the following data points in compression.
    \item Double-check the correctness of the compression based on the given error bound against possible machine epsilon error (line 7-8): specifically, the decompressed value would be reconstructed based on the quantization bin and compared with the true value.
\end{enumerate}

In the following, we analyze the fault tolerance ability of the key procedure of compression upon a computation error occurring in the code segment presented in Figure \ref{fig:sz-ft-analysis} (a), based on five possible cases. We note that the necessary condition to obtain correct decompressed output is that a correct decompressed value must be calculated (\textbf{type-1}) or an unpredictable data handling is called (\textbf{type-2}) during compression; and the same data should be reconstructed during decompression (\textbf{type-3}), which will be used later.
    
\emph{Case 1 - a computation error happens to line 2.} In this case, we need to take into account two possible situations in terms of the deviation of the predicted value affected by the error. 

\begin{itemize}[wide=0pt]
    \item \textit{Situation 1:} the predicted value is changed by the error significantly such that the quantization bin calculated later on falls outside the maximum quantization range (i.e., bin $<$ bin\_max does not hold). In this situation (zone A in Figure \ref{fig:sz-ft-analysis} (b)), the decompressed data will still respect the error bound for sure because of the type-2 behavior. 
    \item \textit{Situation 2:} the impact of the SDC error on the predicted value is relatively small such that the quantization bin is  within the maximum quantization range (i.e., bin $<$ bin\_max still holds). This may cause a significant error to the decompressed data (zone B, C in Figure \ref{fig:sz-ft-analysis} (b)) because of violation of type-3 behavior. The reason is described as follows. On the one hand, the double-checking step (line 7) cannot detect such an error because it would decompress the data point based on the ``wrong'' predicted value such that the reconstructed value will still respect the error bound. On the other hand, it is unlikely that such an SDC error would happen again during the decompression, so that SZ would get a different predicted value for the current data point in the course of decompression and thus a wrong decompressed value on this data point (violation of type-3 behavior). What is even worse is that this decompressed value would also be used to predict other data points in the decompression, such that the errors would be propagated throughout the whole dataset.
\end{itemize}

\emph{Case 2 - A computation error happens to line 3 or 4}. These two lines are naturally resilient due to the type-2 behavior. The unpredicatable data compression is always called (line 10 for zone A and line 8 for zone B), no matter how much the calculated quantization bin deviates (zone B or zone A), 

\emph{Case 3 - A computation error happens to line 6.} This may affect correctness of the decompression data, which will be analyzed based on two possible situations. 
\begin{itemize}[wide=0pt]
\item \textit{Situation 1:} the decompressed data value is deviated significantly because of the SDC such that the following double-checking (i.e., line 7-8) suggests to use unpredictable compression here. So it is resilient because of type-2 behavior.
\item \textit{Situation 2:} the decompressed data value is changed slightly such that it skips the  double-checking step. In this situation, the skewed (wrong) decompressed data value would be used in the prediction of the succeeding data points, and this would lead to the inconsistent prediction results between the compression and decompression. Thus it is not resilient in this situation because of violation of type-3 behavior.
\end{itemize}
  
\emph{Case 4 - A computation error happens to line 7.} Line 7 has very good resilience but not perfect. Obviously, if line 7 makes a false result to be true, it is resilient because of the unpredictable data solution (type-2 behavior). If line 7 makes a true result to be false, it is not resilient because of the impact of machine epsilon. However, in our fault tolerant design, we do not protect this part because the likelihood of this situation is extremely small. This situation happens only when the original real value is located right on the edge of a quantization bin. To be more specific, a test shows only 24 out of $512^3$ data points (NYX dataset, relative error bound 1E-3) will make line 7 true.

\subsubsection{SDC resilience in lossless compression}
We will show our solutions are able to detect SDCs that occur in lossless compression in Section~\ref{sec:ft-decompression}.

All in all, in terms of the SZ lossy compression framework, the only concern regarding fault tolerance during the compression procedure is on the correctness of the predicted value (i.e., line 2 in Figure \ref{fig:sz-ft-analysis} (a)) and the correctness of data decompression during the compression (i.e., line 6). To address this issue, we developed an efficient selective instruction duplication method, to be described in Section \ref{sec:methodology} in detail.

\subsection{SDC Resilience -- Memory error}

Now, we analyze the resilience against the memory errors occurring in different places, such as input data, regression coefficients and quantization bin index array, respectively. 

\subsubsection{SDC resilience against memory error in inputs}
\label{sec: mem err in input}
Since the input data (i.e., original data) occupies the significant portion of the memory footprint, we have to protect it against potential SDC errors. The input data is used in the following steps: 1. computing the regression coefficients; 2. sampling and estimating the compression error of both regression and Lorenzo predictor; 3. data prediction and calculation of the difference between predicted data and original data and handling unpredictable data. We find that: for the first two steps, similar to the analysis in Section~\ref{sec:analysis of comp err in reg coeff},  the memory error in input data will only impact the compression ratio and keep the correctness of decompressed data. However, step 3 must use genuine uncorrected input data since that is where the compression happens. With a corrupted input in step 3, the decompressed data will be calculated based on that corrupted value which is obviously SDC prone.

We will leverage the above finding to reduce the overhead of checksum calculations since it discloses the fact that the corrupted values may not affect the correctness of decompressed data in the first 2 steps (i.e., error detection/correction for those parts are not necessary).

\subsubsection{SDC resilience against the memory error in regression coefficients}
The memory usage of regression coefficients are found to be very small compared to the overall memory usage such that this part does not need particular protection. Each data block will maintain at most 4 coefficients (for 3D dataset). Thus, the coefficients only take $\frac{4}{block size}$ of the overall memory. For a 3D example, usually the block size is 8X8X8 which means the coefficients take only $\frac{1}{128}$ of overall memory. 


\subsubsection{SDC resilience against the memory error in quantization bin index array} In SZ, the quantization bin index array (to be called bin array for simplicity) is an array used to record how much the predicted value deviates from the original value for each data point. The element in the array is a positive integer if the data is predictable; otherwise, the element is 0, indicating that the data needs to be compressed/decompressed by unpredictable compression method. Obviously, if the bin array is corrupted by some memory error, the decompressed data will not be correct. So, the array is not resilient to memory error. Also, since the prediction is a critical stage that contributes the portion of the overall execution time, the likelihood of error happening during this stage is higher than other stages, thus we have to protect the bin array in this stage. Specifically, we carry out two different checksums on each block right after all the data inside the block are processed, such that we are able to detect and correct the possible corrupted data by double-checking the checksum values later on (e.g., during the Huffman encoding stage).

%% file: tex/methodology.tex
\section{Error Tolerance Methodology}
\label{sec:methodology}
Our SDC resilient SZ design is done in three aspects.
First, we eliminate the data dependency between adjacent blocks; second, we use selective instruction duplication to ensure correct computation; third, we use checksums to detect and correct corrupted values caused by memory errors. 

\subsection{Blockwise independent design}

In the following, we discuss how to eliminate the dependency between blocks, such that any SDC error can be confined within a small block, improving the robustness significantly. 

The key difference between the original SZ and our independent-block based compression is that we now treat each block of data as separately with each other. Specifically, we apply the prediction and quantization inside each block individually and make sure the compressed data of one block is totally independent with others'. This requires many changes to the original SZ development. For instance, we need to record the compressed size of each block after we finish the compression for that block. 
Both recording the bin array and Huffman encoding need to be done individually per block.

Another significant advantage in the independent-block based compression design is that one can perform random-access decompression efficiently by specifying a specific region in space. To this end, we implement random-access support in our implementation, such that the decompression speed can be improved significantly if the user just wants to decompress a small region in the whole dataset. The corresponding experimental results will be presented in Section \ref{sec:experiment}. Moreover, such an independent-block based compression also makes the parallelism of SZ much easier to port on many-core architectures, such as GPU.

\subsection{Fault tolerant compression}
\begin{algorithm}
\caption{\textsc{Soft Error Resilient SZ Compression}} \label{alg: ft sz cmpr}
\renewcommand{\algorithmiccomment}[1]{/*#1*/}
\footnotesize
\begin{flushleft}
\textbf{Input}: original input data (denote by $ori[]$), user defined error bound (denoted by $e$).\\
\textbf{Output}: compressed data in byte \textcolor{blue}{and compressed $sum$ of blocked decompressed data}
\end{flushleft}

\begin{algorithmic}[1]

\FOR {each block (block $i$) of the input data}

\STATE Compute the regression coefficients
\STATE \textcolor{blue}{Get $sum_{in}$[$i$] on input by Equation~(\ref{equa:sum})} \COMMENT{\textcolor{blue}{for SDC in input data}}
\STATE \textcolor{blue}{Get $isum_{in}$[$i$] on input by Equation~(\ref{equa:isum})} \COMMENT{\textcolor{blue}{for SDC in input data}}
\ENDFOR

\FOR {each block (block $i$) of input data}
\STATE Sample and estimate $E_{reg}$ and $E_{lor}$ 
\STATE $indicator[i] \gets $ the one with smaller error \COMMENT{regression or lorenzo}
\ENDFOR

\FOR {each block (block $i$) of input data}
\STATE \textcolor{blue}{Do memory error detection and correction using $sum_{in}$ and $isum_{in}$}

\IF{$indicator[i] == regression $}
\STATE $f() \gets regression\  predictor$
\ELSE
\STATE $f() \gets lorenzo\  predictor$
\ENDIF

\FOR {each data point, $ori$, in the data block }

\STATE\textcolor{blue}{ $pred' \gets f_{dup}()$ \COMMENT{\textcolor{blue}{$f_{dup}$(): instruction duplicated  $f()$}}}
\STATE \textit{diff} $\gets ori - pred'$
\STATE $q\_bin$ $\gets$ $quant$(\textit{diff},$e$) \COMMENT{get quantiz. bin based on \textit{diff},$e$}
\IF {$q\_bin$ is not in the acceptable bin range}
\STATE Compress $ori$ as unpredictable
\ELSE
\STATE \textcolor{blue}{ Calculate $sum_{q}$, $isum_{q}$ for $q\_bin[]$}
\STATE{\textcolor{blue}{ $dcmp \gets dec_{dup}(q\_bin, pred') $ \COMMENT{$dec_{dup}$() is instruction duplication based version of $dec()$}}}
\IF { $|ori - dcmp| > e$}
\STATE Compress $ori$ as unpredictable
\ENDIF
\STATE \textcolor{blue}{$sum_{dc}$[$i$]+=\emph{dcmp} \COMMENT{cksum for decompressed data of block $i$ }}
\ENDIF

\ENDFOR
\ENDFOR


\STATE Construct Huffman tree
\FOR {each block of $q\_bin[]$}
\STATE \textcolor{blue}{Do memory error detection and correction using $sum_q$ and $isum_q$}
\STATE Encode $q\_bin[]$ by Huffman tree
\ENDFOR
\STATE Compress encoded $q\_bin$ by lossless method (Zstd)
\STATE Write compressed $q\_bin$ and unpredictable data to byte file
\STATE \textcolor{blue}{ Compress $sum_{dc}$[] by lossless method (Zstd) and write to file}
\end{algorithmic}

\end{algorithm}

We present our SDC resilient compression method in Algorithm~\ref{alg: ft sz cmpr}. We highlight the lines related to our fault tolerance design in blue font. Line 3 and 4 are calculating checksums for input data, in order to detect possible SDC errors striking the input data later on. 
As we discussed in Section~\ref{sec: mem err in input}, we do not need to detect memory error in the input data during computations for regression coefficients and compression error estimation. We only detect whether the input data encounters memory errors before the data prediction gets started (line 11). If a data corruption is detected (by $sum_{in}$), it can be located and recovered by the pair of checksums (i.e., $sum_{in}$ and $isum_{in}$) applied on input data. Then, we protect the quantization bin array against memory errors (line 24 and 35). Line 29 and 40 are designed for detecting possible SDC errors occurring in the decompression stage, to be detailed later. For the computation errors, instruction duplication can be used. According to our analysis in Section~\ref{sec: comp err rslc}, only data prediction (line 18) and calculating decompressed data (line 25) need to be protected by instruction duplication.

\subsection{Fault tolerant decompression}
\label{sec:ft-decompression}
The SDC resilient SZ decompression is presented in Algorithm \ref{alg: ft sz decmpr}. Line 1-9 refers to the regular block-wise data decompression of SZ. Our resilience design starts from line 10.  We constructed the checksums for each block and compressed the checksum array (i.e., $sum_{dc}$[]) by lossless compression (Zstd) during the data compression. Accordingly, we need to decompress $sum_{dc}$ (line 10) before the error detection. Our idea is leveraging such checksums of decompressed data (i.e., $sum_{dc}$[]) constructed during the compression to detect possible errors that happen during the decompression. Specifically, after performing the data decompression for each block (line 1-9), our algorithm will calculate the corresponding checksums for each block of decompressed data and compare the checksums to $sum_{dc}$[] (line 12-13). If they are not consistent, some errors must happen during the decompression. So, the algorithm will decompress this block by random-access decompression (line 14), meaning the compressed bytes are reloaded. If the checksum is consistent, we know some memory or computation error is detected (line 17). If inconsistent the second time, we can conclude that the SDC error likely happens during the lossless compression, which will be reported to users (line 19). 

\begin{algorithm}
\caption{\textsc{Soft Error Resilient SZ Decompression}} \label{alg: ft sz decmpr}
\renewcommand{\algorithmiccomment}[1]{/*#1*/}
\footnotesize
\begin{flushleft}
\textbf{Input}: The SZ compressed file in byte (cmp\_data) \textcolor{blue}{and compressed $sum$ for blocked decompressed data ($sum_{dc}[]$)}. \\
\textbf{Output}: Decompressed data with bounded error compared to original data.
\end{flushleft}
\begin{algorithmic}[1]

\STATE Decompress cmp\_data by lossless compressor (Zstd)
\FOR {each block}
\STATE $q\_bin[] \gets$ decode using Huffman tree
\IF {it was compressed by Lorenzo}
\STATE $dec\_data[] \gets $ Lorenzo decompression
\ELSE
\STATE $dec\_data[] \gets $ regression decompression
\ENDIF
\ENDFOR
\STATE \textcolor{blue}{Decompress $sum_{dc}[]$ by lossless compressor (Zstd)}
\FOR {\textcolor{blue}{each block of decompressed data (block index = $i$)}}
\STATE \textcolor{blue}{ Calculate checksum (denoted $sum_i$) for this block of $dec\_data[]$}
\IF {\textcolor{blue}{ $sum_i$ $\neq$ $sum_{dc}[i]$}}
\STATE \textcolor{blue}{Reexecute line 4-9 for this block} \textcolor{blue}{\COMMENT{random-access decompression}}
\STATE \textcolor{blue}{ Calculate checksum (denoted $sum_i$) for this block of $dec\_data[]$}
\IF {\textcolor{blue}{ $sum_i$ = $sum_{dc}[i]$}}
\STATE \textcolor{blue}{Report: memory/computation error detected but corrected}
\ELSE
\STATE \textcolor{blue}{Report: SDC in compression; Return}
\ENDIF
\ENDIF
\ENDFOR
\end{algorithmic}

\end{algorithm}


\vspace{-2mm}
\subsection{Avoiding round off errors in checksums}
Since the input data and the decompressed data are both floating point numbers, round off errors in the checksums may introduce inaccurate memory error corrections. To avoid the impact of round off error, we treat the floating point numbers as unsigned 32-bit integers and then calculate checksums based on these integers. We first describe how the checksum is performed on the 32-bit single-precision floating point data as an example and then discuss how to extend it to 64-bit double-precision floating point values.

Given a data block of 32-bit floating point values, for each element, we put all its 32 bits in a temporary variable and treat the bits in that variable as a 64-bit unsigned integer with the first 32 bits being flushed to 0. We then add that integer to the checksum which is also a 64-bit unsigned integer. Finally, we get the checksum represented by a 64-bit unsigned integer for this data block. Notice that the ``checksum'' here is not equal or approximate to the real sum of the data block because it is calculated based on integer interpretation of the bits instead of floating point. Thus, it is immune to NaN/Inf issues that happens only to floating point numbers. Using the 64-bit unsigned integer representation, we can have the checksum hold up to $(2^{32}+1)$ 32-bit unsigned integers without overflow because the maximum 64-bit unsigned integer $(2^{64}-1)$ divided by maximum 32-bit unsigned integer $(2^{32}-1)$ is equal to $(2^{32}+1)$. That is fairly enough to totally avoid the overflow since each data block in SZ has only 1000 data points (such as 10$\times$10$\times$10 block) in general. With all these techniques, we can provide bit-level error detection and correction. The main difference from Demmel's work \citep{reproduciblesum} is that we are actually doing integer-based summation instead of the sum based on floating point numbers. 

To extend to 64-bit double precision  numbers, we just need to treat each double value as two 32-bit unsigned integers. So it is reduced to the above case.

\vspace{-3mm}
\subsection{Impact to compression ratio without protecting regression and sampling}
\label{sec:no reg sampling protct impact}
As mentioned previously, we do not protect the computation in regression and sampling in that the errors during this period would not affect the correctness of decompressed data and just have tiny impact to the compression ratios. In what follows, we derive theoretically the upper bound of the compression ratio decrease affected by the computation errors happening during the regression or sampling. We denote the compression ratio of SZ in error free run by $R_0$; the number of data blocks by $n$. For simplicity, we assume that the compression ratio for each block is identical with each other. In the worst case, the error in regression or sampling will at most reduce the compression ratio to be 1, which means that it does not reduce the size of that block of data. Consequently, we can derive the maximum compression ratio decrease as \emph{CR}\_\emph{decrease} = ($\frac{R_0-1}{R_0+n-1}$)$\times$100\%.
Based on the above equation, the upper bound of compression ratio decrease depends on the error free compression ratio and the block size. For example, if the block size is set to 6X6X6 and the compression ratio is 10, and if the input data is around 864 MB, then there will be $10^6$ data blocks. The compression ratio decrease would be bounded within $\frac{10-1}{10-1+10^6} < 0.1\%$, which is negligible to the overall compression ratios.

%% file: tex/evaluation.tex
\section{Experimental Evaluation}
\label{sec:experiment}


\subsection{Experimental Setup}

In this subsection, we describe how we set the experiments in our evaluation, including applications, error injections, and experimental environment. 

\subsubsection{Applications}
\begin{figure}[ht]
\centering
\subfigure[Original image]{
{\includegraphics[width=4cm,bb=0 0 514 512]{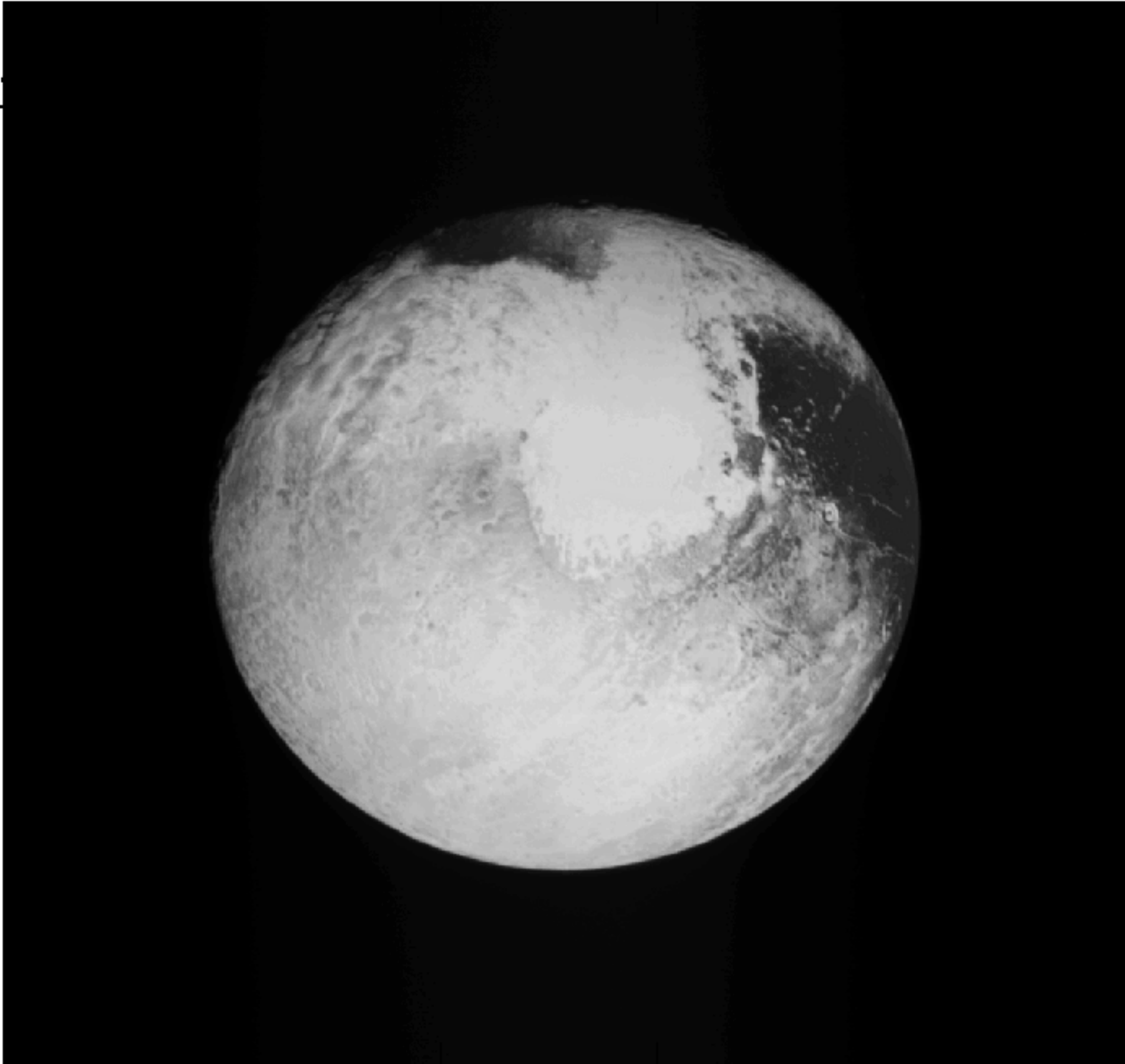}}}
\subfigure[SZ decompressed image]{
{\includegraphics[width=4cm,bb=0 0 514 512]{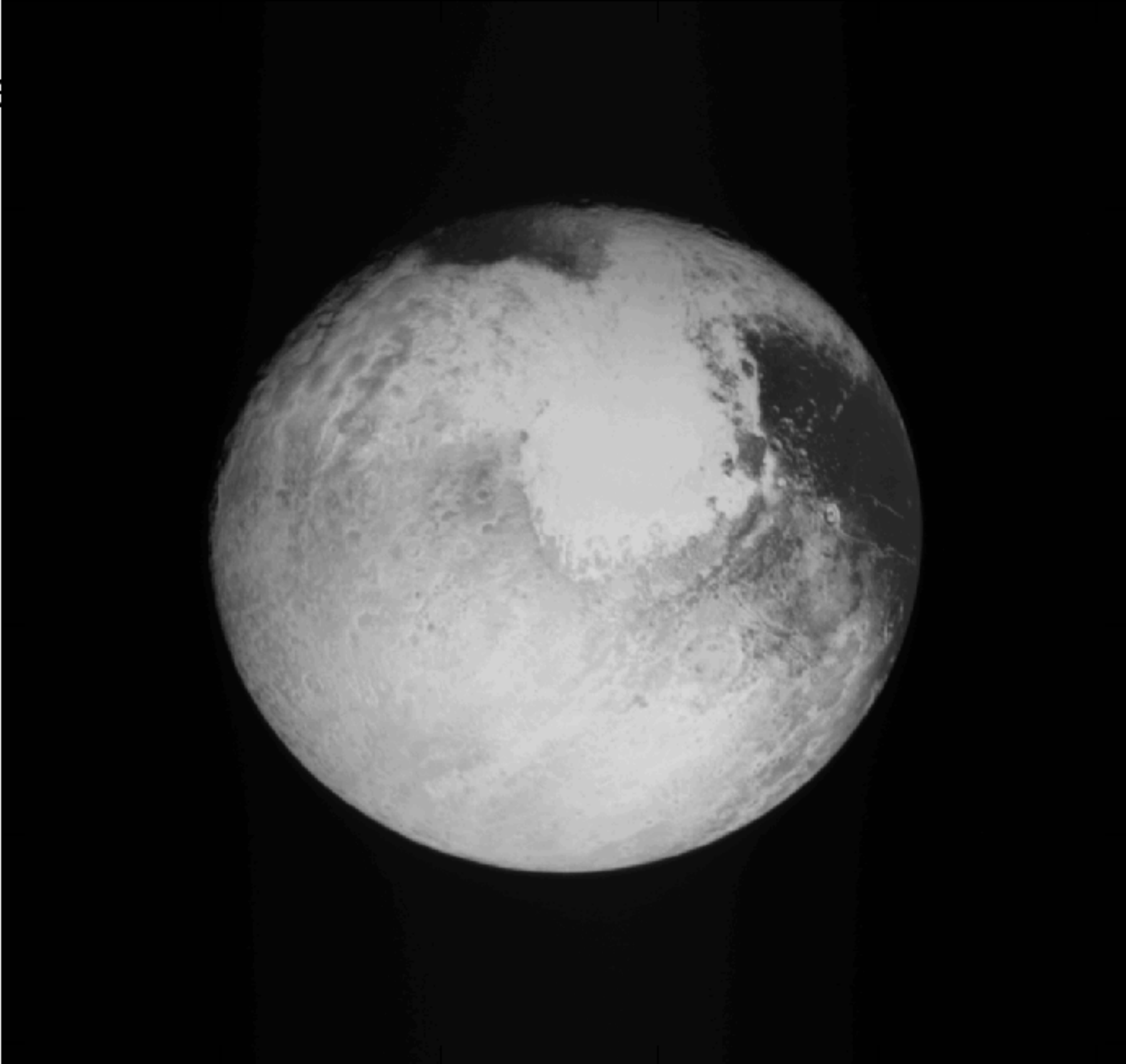}}}
\caption{Visualization of Original Data vs. Decompressed Data (Pluto photo taken by New Horizons \citep{new-horizons}; SZ compression using Value-range based error bound: 1E-3)}
\label{fig:pluto}
\end{figure}

We evaluate our SDC resilient error-bounded SZ compressor on three real scientific datasets: NYX, Hurricane, and SCALE-LETKF (SL for short). We also evaluate our fault tolerant compressor using 20 Pluto images provided by Plantary Data System (PDS) \citep{pds}. Those images were taken by New Horizons space probe \citep{new-horizons} in aerospace which is an error-prone environment because of potential impact of cosmic rays. The description to these datasets is presented in Table~\ref{tab:dataset information}. For the Pluto image data, we perform the error-bounded lossy compression such that the visual quality can be maintained very well, as illustrated in Figure~\ref{fig:pluto}.
\begin{table}[]
    \centering
    \caption{Basic dataset information}    
    \vspace{-.3cm}
    \footnotesize
    \begin{tabular}{|c|c|c|c|}
    \hline
    \textbf{Dataset}&\textbf{\# Fields} & \textbf{Dimensions} & \textbf{Science}\\
    \hline
    NYX&6&512X512X512& Cosmology\\
    \hline
    Hurricane &13&100X500X500&Climate\\
    \hline
    SCALE-LETKF (SL) & 6 & 98X1200X1200 & Weather\\
    \hline
    NASA: Pluto & 1 &1028X1024&Aerospace\\
    \hline
    \end{tabular}
    \label{tab:dataset information}
\end{table}

\subsubsection{Error injections with two modes}

\paragraph{Evaluation mode A - source-code level error injection. } Like most ABFT work \citep{xinfft,jieyangipdps}, we inject errors at the source code level and only inject errors to the main data structures. Specifically, as for the memory errors in input data and quantization bin array, we randomly choose an index from the array and randomly flip a bit of the selected data value during the compression. Thus, we simulate memory error randomness both in time and location.
We inject them after the checksums are applied on input data (i.e., $sum_{in}$[] and $isum_{in}$[]).  
To simulate the computation errors when calculating regression coefficients, sampling and estimating compression error of Lorenzo and regression, we randomly select a data point in a random block and then change its value by injecting a random bitflip error. We exclude the evaluation of computation errors in prediction as it is already protected by instruction redundancy. 
\vspace{-2mm}
\paragraph{Evaluation mode B - system level error injection.}
Besides the evaluation mode A (memory errors happens only to the data we protected), we also follow a Checkpoint-based Fault Injection (CFI) \citep{cfi} model to inject random error(s) to the whole memory consumed during the compression. We adopt a system-level checkpointing toolkit - Berkeley Lab Checkpoint/Restart (BLCR) \citep{blcr}, which can dump the whole memory of a running process to disk as a checkpoint and then restart its execution from that checkpoint. In our experiment, we select a random time stamp during the whole compression period. Then, we set a checkpoint by saving the whole memory at that time stamp using BLCR and kill the process. We then inject a random bitflip error in the checkpoint file and restart the process by the bit-flipped checkpoint. We inject 1, 2 or 3 errors and perform 500 runs per test for both fault tolerant SZ and original unprotected SZ.

\vspace{-2mm}
\subsubsection{Experimental Environment}
We run experiments on a supercomputer \citep{bebop}. Inside each computing node are two Intel Xeon E5-2695 v4 processors totalling 36 cores. The POSIX I/O \citep{posixio} with mode, file-per-process, is used for parallel data reading and writing. We implement our solution in SZ's source code and call it ftrsz (or FT-SZ) in the following text. We alter the order of value additions in the duplicated computation of data prediction, which can effectively prevent the compiler from overlooking this operation, and the execution time overhead can thus be measured correctly. 

\vspace{-2mm}
\subsection{Evaluation of Independent-block Compression}

We first evaluate our designed independent-block based SZ compression (a.k.a., random-access based compression).

\subsubsection{Exploration of The Best Block Size}

It is important to determine an appropriate block size in our independent-block based compression framework.
We determine the best block size by a comprehensive analysis in terms of rate-distortion with masses of experiments using different block sizes, as the optimal block size is hard to find for different datasets by theory.

\begin{figure}[ht]
\centering
\hspace{-8mm}
\subfigure[NYX velocity\_x]{
\raisebox{-1cm}{\includegraphics[scale=0.38]{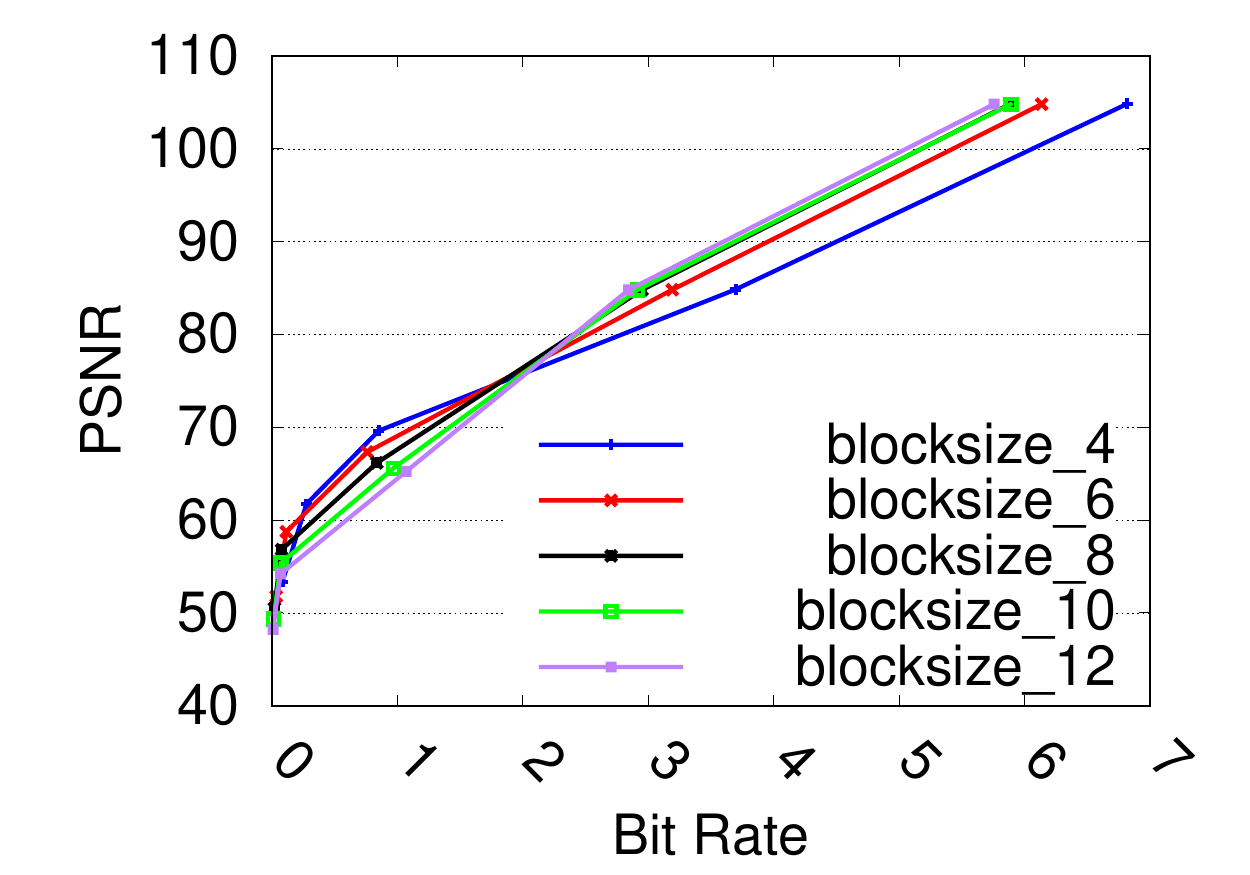}}}
\hspace{-8mm}
\subfigure[Hurricane TCf48]{
\raisebox{-1cm}{\includegraphics[scale=0.38]{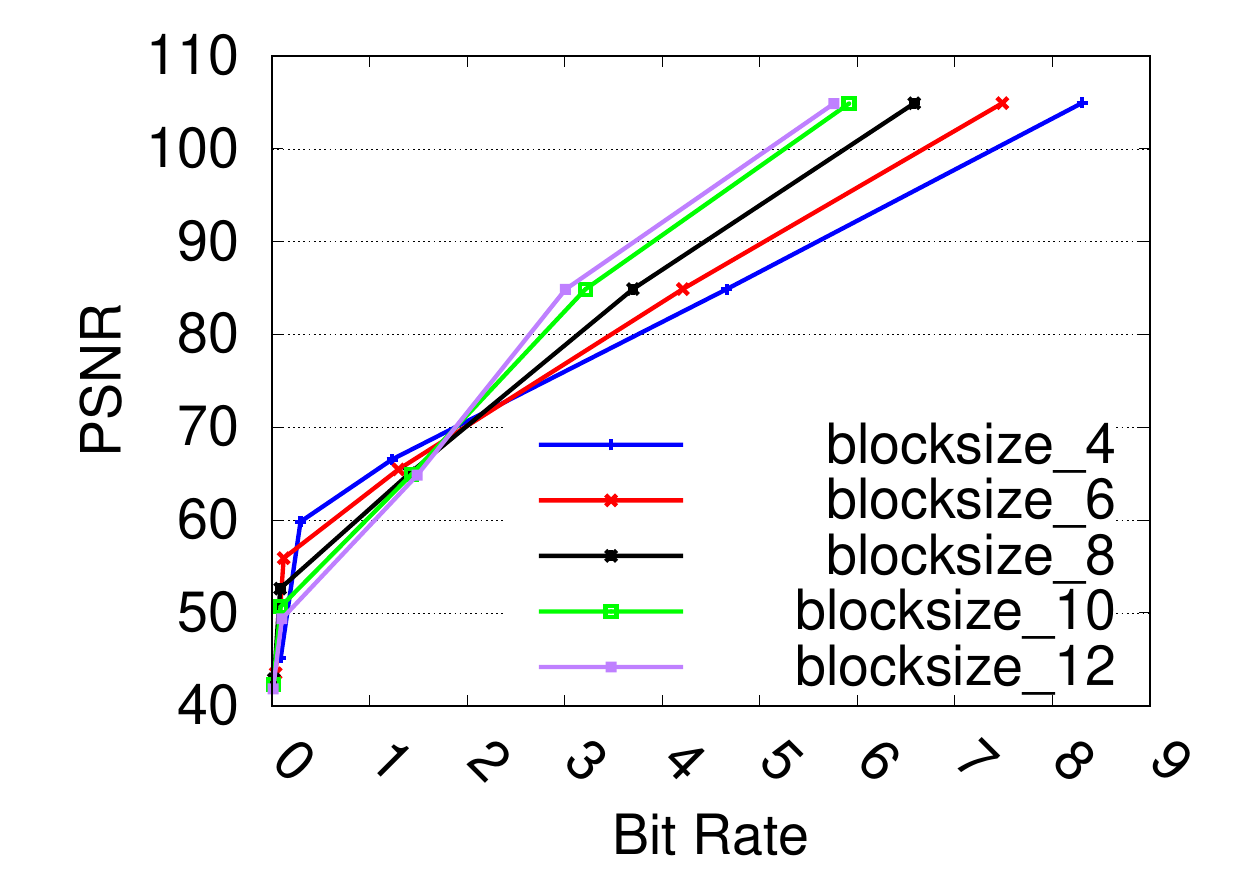}}}
\hspace{-8mm}
\caption{Rate distortion with different block sizes}
\label{fig: block_size}
\end{figure}

We evaluate the compression results using the block size of 4x4x4 through 20x20x20. We exemplify the rate-distortion with cosmological NYX simulation data (velocity\_x field) and climate hurricane simulation data (TCf48 field) with five different block sizes in Figure~\ref{fig: block_size}. As shown in the figure, small block sizes (such as 4x4x4 and 6x6x6) may lead to high PSNR in the cases with low bit-rates (such as $\leq$2); large block sizes (such as 8x8x8 $\sim$ 12x12x12) would be clearly better than the small block sizes on high bit-rates. The reason is explained as follows. 
For the over-small block sizes such as 4x4x4, the overhead of storing the regression-coefficients appears relatively high compared to the overall compressed size. For the over-large block sizes such as 20x20x20, the linear-regression based predictor cannot get a good fitting for the data. Based on our experiments with multiple simulation data, we set the block size to 10x10x10 in our implementation because it has much better compression ratios (i.e., low bit-rate) in the hard-to-compress cases than other block sizes, while it exhibits comparative compression ratios with other block sizes in the cases with relatively low bit-rates.

\subsubsection{Evaluating independent-block decompression}
\begin{figure}
    \centering
    \includegraphics[scale=0.34]{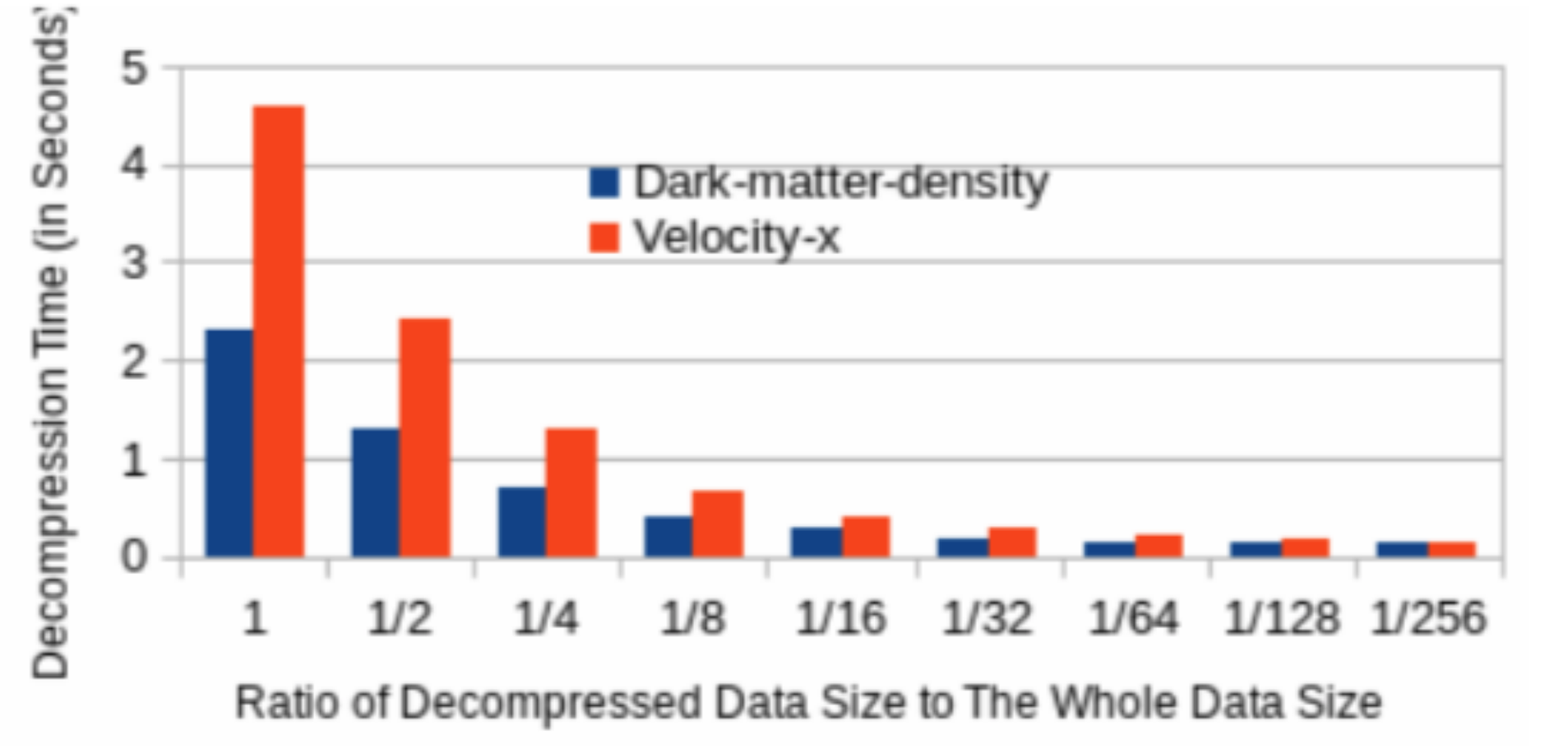}
    \caption{Efficiency of random access decompression}
    \label{fig:rand effi decmp}
\end{figure}

The biggest advantage of the independent-block based implementation is very fast decompression speed if the users just want to extract a small sub-block of data.
Moreover, as we discussed in Section \ref{sec:ft-decompression}, this design can also help correct the errors very quickly upon a detection of problematic blocks by checksums. In Figure~\ref{fig:rand effi decmp}, we present the decompression times with different data sizes compared to the whole dataset. The x-axis indicates the ratio of the decompressed data size to the whole data size. 
In the figure, we observe that the decompression time decreases approximately linearly with decreasing data size in the decompression, which confirms the high efficiency of random-access decompression.

\subsection{Error free experimental results}

One key indicator is how much overhead (including compression ratio overhead and execution time overhead) would be introduced by the SDC detection in the compressor.

\begin{table}[]
    \centering
    \caption{Compression ratio degradation of random-access SZ (rsz) and fault-tolerant random-access SZ (ftrsz)}
    \vspace{-2mm}
\resizebox{0.99\columnwidth}{!}{    
    \begin{tabular}{|c|c|c|c|c|c|c|c|c|}
    \hline
error bound: & 1E-3 &1E-4 &1E-5 &1E-6 &  1E-3 &1E-4 &1E-5 &1E-6 \\
\hline
\hline
\multicolumn{1}{|c}{} & \multicolumn{4}{|c}{\textbf{NYX}} & \multicolumn{4}{|c|}{\textbf{Hurricane}}  \\
\hline
sz: &17.0& 7.7&  4.6&  3.1 &8.4 & 5.1 & 3.1 & 2.4\\
\hline
rsz decrease:&8.7\%& 3.7\%& 3.1\% &3.2\%& 8.5\% & 4.7\% & 1.2\% & 1.5\% \\
\hline
ftrsz decrease:&10.7\% & 4.7\% & 3.7\% & 3.6\%& 9.3\% & 5.2\% & 1.6\% & 1.7\%\\
\hline
\hline
\multicolumn{1}{|c}{}&\multicolumn{4}{|c}{\textbf{SCALE-LETKF (SL)}}&\multicolumn{4}{|c|}{\textbf{Pluto}}\\
\hline
sz: &19.1 & 8.7 & 5.2 & 3.7 &7.1 & 4.0 & 3.4 & 3.2\\
\hline
rsz decrease:& 23.6\% & 21.3\% & 13.5\% & 9.1\%& 4.2\% & 0.3\% & 0.1\% & 0\%  \\
\hline
ftrsz decrease:& 24.9\% & 21.9\% & 13.9\% & 9.4\%& 5.6\% & 0.8\% & 0.1\% & 0\%\\
\hline
    \end{tabular}}
    \label{tab:all cmp ratio}
\end{table}

\begin{table*}
    \centering
    \caption{Percentage of runs whose maximum absolute error is within error bounds in sz and ftrsz}
    \vspace{-2mm}
    \begin{adjustbox}{max width=1\textwidth}
    \begin{tabular}{|p{2cm}|p{1.3cm}|p{1.3cm}|p{1.3cm}|p{1.3cm}|p{1.35cm}|p{1.35cm}|p{1.35cm}|p{1.35cm}|p{1.4cm}|p{1.4cm}|p{1.4cm}|p{1.4cm}|}
    \hline
       \multicolumn{1}{|c}{} & \multicolumn{4}{|c|}{\textbf{injecting errors in input data}} & \multicolumn{8}{c|}{\textbf{injecting errors in quantization bin array}} \\
    \hline
       \multicolumn{1}{|c}{} & \multicolumn{4}{|c|}{Successful runs with correct decompressed data} & \multicolumn{4}{c|}{Successful runs without correct decompressed data} & \multicolumn{4}{c|}{Normal runs without core-dump segmentation faults} \\
    \hline
    error bounds: & 1E-3 & 1E-4 & 1E-5 & 1E-6 & 1E-3 & 1E-4 & 1E-5 & 1E-6 & 1E-3 & 1E-4 & 1E-5 & 1E-6\\
    \hline
    sz &60\%&57\%&49\%&48\% & 3\% & 1\% & 1\% & 0\% & 34\% & 34\% & 49\% & 54\%\\
    \hline
    ftrsz &100\%&100\%&100\%&100\%&100\%&100\%&100\%&100\%&100\%&100\%&100\%&100\%\\
    \hline
    \end{tabular}
    \end{adjustbox}
    \label{tab:input/type mem err}
\end{table*}

\subsubsection{Compression ratio overhead} 
Since we store the checksum $sum_{dc}$[] during the compression in order to verify the correctness of the decompressed data, the compression ratio could be degraded more or less. Table~\ref{tab:all cmp ratio} presents the compression ratios of the original SZ (denoted as sz) and the relative decreases of compression ratios under the independent-block based SZ (or random-based SZ, abbreviated as \textit{rsz}) and fault-tolerant random-access SZ (denoted as ftrsz), respectively. It is observed that our proposed solution incurs only 0$\sim$10.7\% degradation on compression ratio for NYX, Hurricane and Pluto data, and the degradation level decreases with decreasing error bounds. The SL dataset exhibits 9.4$\sim$24.9\% compression ratio degradation, which mainly comes from the overhead introduced by the random-access design. 

\vspace{-2mm}
\subsubsection{Execution time overhead}

\begin{figure}
    \centering
    \hspace{-8mm}
    \subfigure[Compression]{\includegraphics[width=4.5cm]{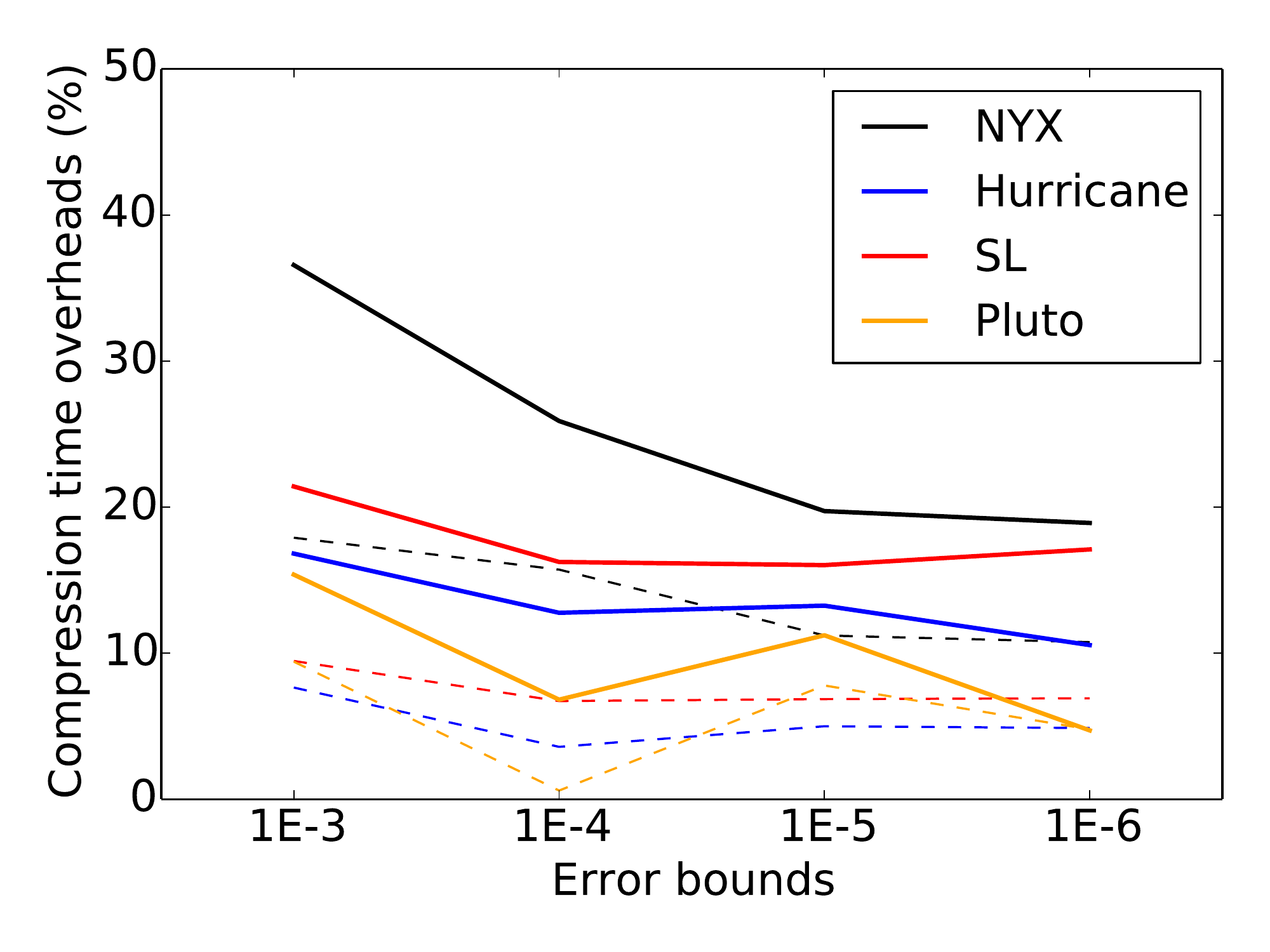}}
    \hspace{-4mm}
    \subfigure[Decompression]{
    {\includegraphics[width=4.5cm]{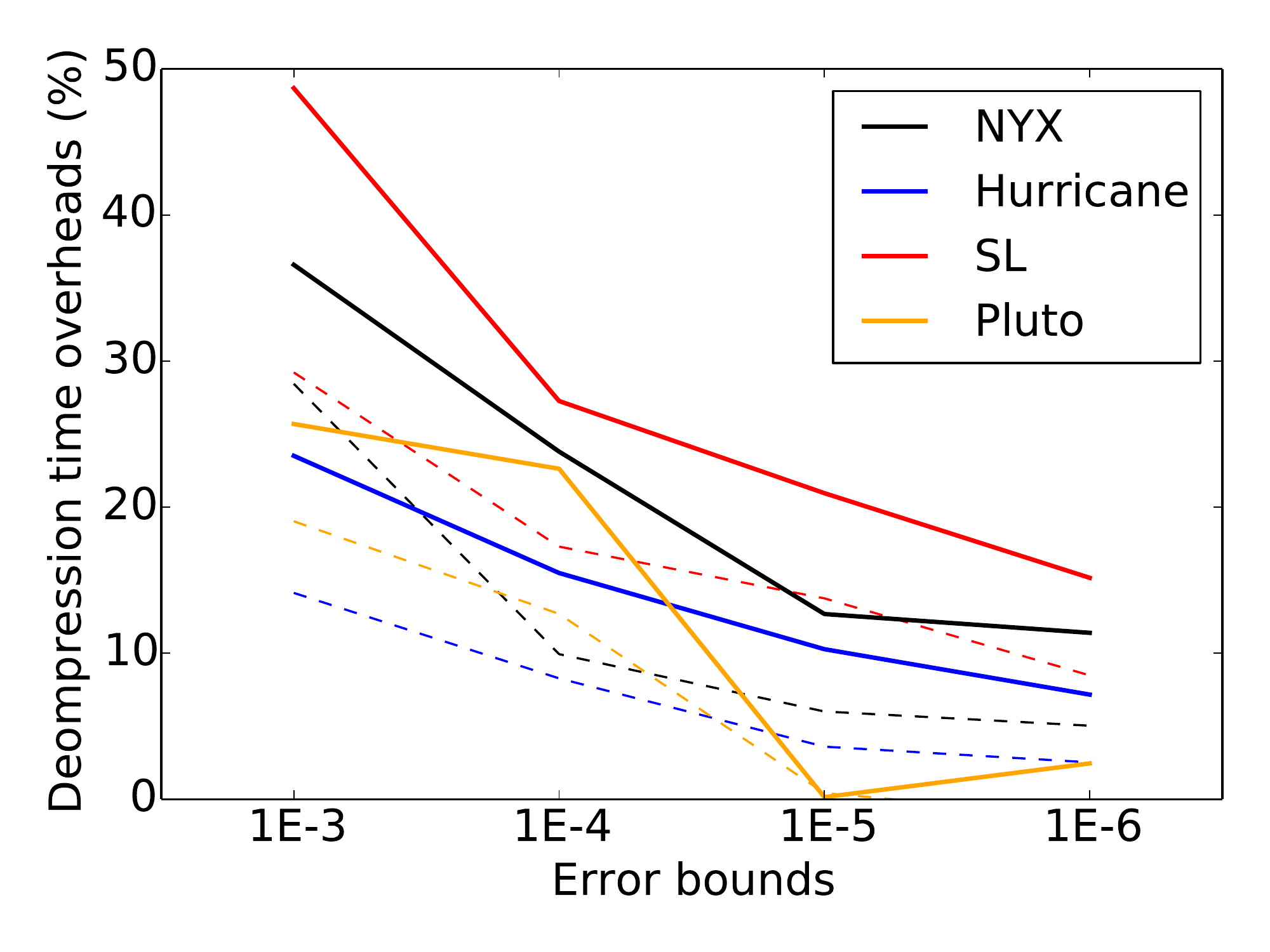}}}
    \hspace{-8mm}
    \caption{Compression time and decompression time overheads. Dash lines are random access SZ; solid lines are fault tolerant random access SZ.}
    \label{fig:cmp dcmp time overheads}
\end{figure}

We evaluate the time overheads introduced by our fault tolerance codes added to SZ when there are no errors. We show the results in both compression and decompression in Figure~\ref{fig:cmp dcmp time overheads}. We can see from Figure~\ref{fig:cmp dcmp time overheads} that in most cases, the rsz and ftrsz incur about 5$\sim$20\% overheads in compression time and 2$\sim$30\% overheads in decompression time. Such time overhead, actually, are negligible compared to the total I/O time on a PFS because of potential I/O bottleneck, which will be demonstrated in the end of this section.

\subsection{Error injected experimental results}

\subsubsection{Resilience against memory errors in input and quantization bin array (evaluation mode A)}

We first inject memory errors into the input array and bin array to verify that our proposed solution can still ensure the decompressed data within the user defined error bounds. 

In this experiment, we observe that various fields exhibit similar results. As such, we present the results based on the field of dark matter density in NYX dataset as an example. For every error bound, we repeat running sz and ftrsz for 100 times, each with randomly injected memory errors in input and quantization bin array. 

As shown in Table~\ref{tab:input/type mem err}, our proposed fault tolerance solution can always yield correct decompressed results when the memory errors are injected in input data or quantization bin array. The 100\% correctness of the decompressed data under ftrsz also means that our solution is immune to the round-off errors. In comparison, for the original SZ without our techniques, we can see that only 48$\sim$60\% runs can yield error bounded decompressed data when the input data experiences memory errors. As the memory error corrupts a value in the bin array, the situation gets worse because some of the memory errors may cause core-dump segmentation fault, which happens in the case that the corrupted values turn out to be a fresh value such that it is beyond the range of the constructed Huffman tree. 
As shown in the right side of Table~\ref{tab:input/type mem err}, under the original SZ compression, only 34$\sim$54\% runs can complete without segmentation faults; and only 0-3\% runs can complete with correct decompressed data. 

As for the extra time overheads introduced by the detection/correction of errors in our fault tolerance method, we conduct error injected experiments for all three datasets. The extra overheads compared to ftrsz in an error-free case are all less than 1\% for any error bound. This is because the case with injected errors only incurs one more block of checksum calculation, which is negligible to the overall execution time.

\vspace{-3mm}
\subsubsection{Resilience against memory errors happening anywhere (evaluation mode B)}

\begin{figure}[h]
\centering
\vspace{-1mm}
\hspace{-8mm}
\subfigure[Runs without crashes]{\includegraphics[scale=0.36]{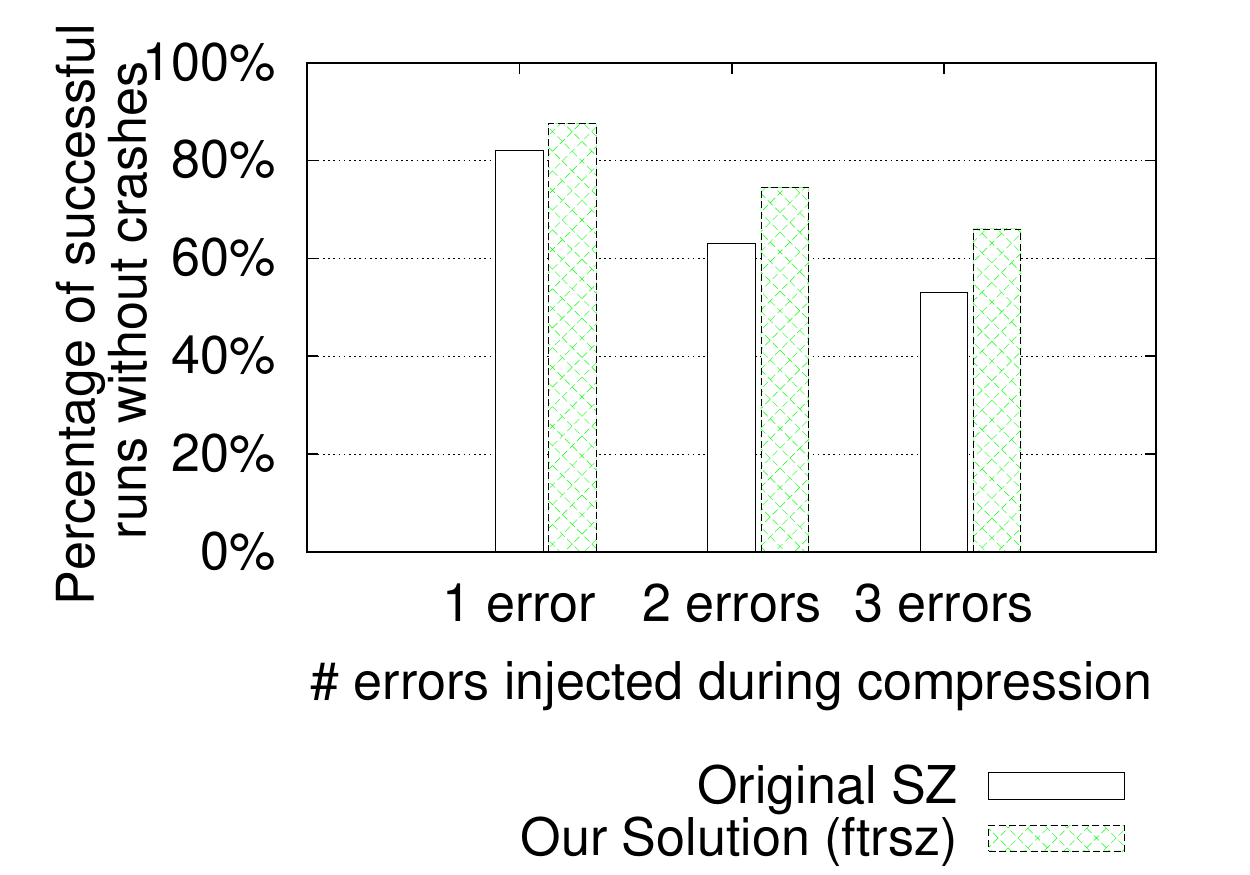}}
\hspace{-6mm}
\subfigure[Runs without SDC]{\includegraphics[scale=0.36]{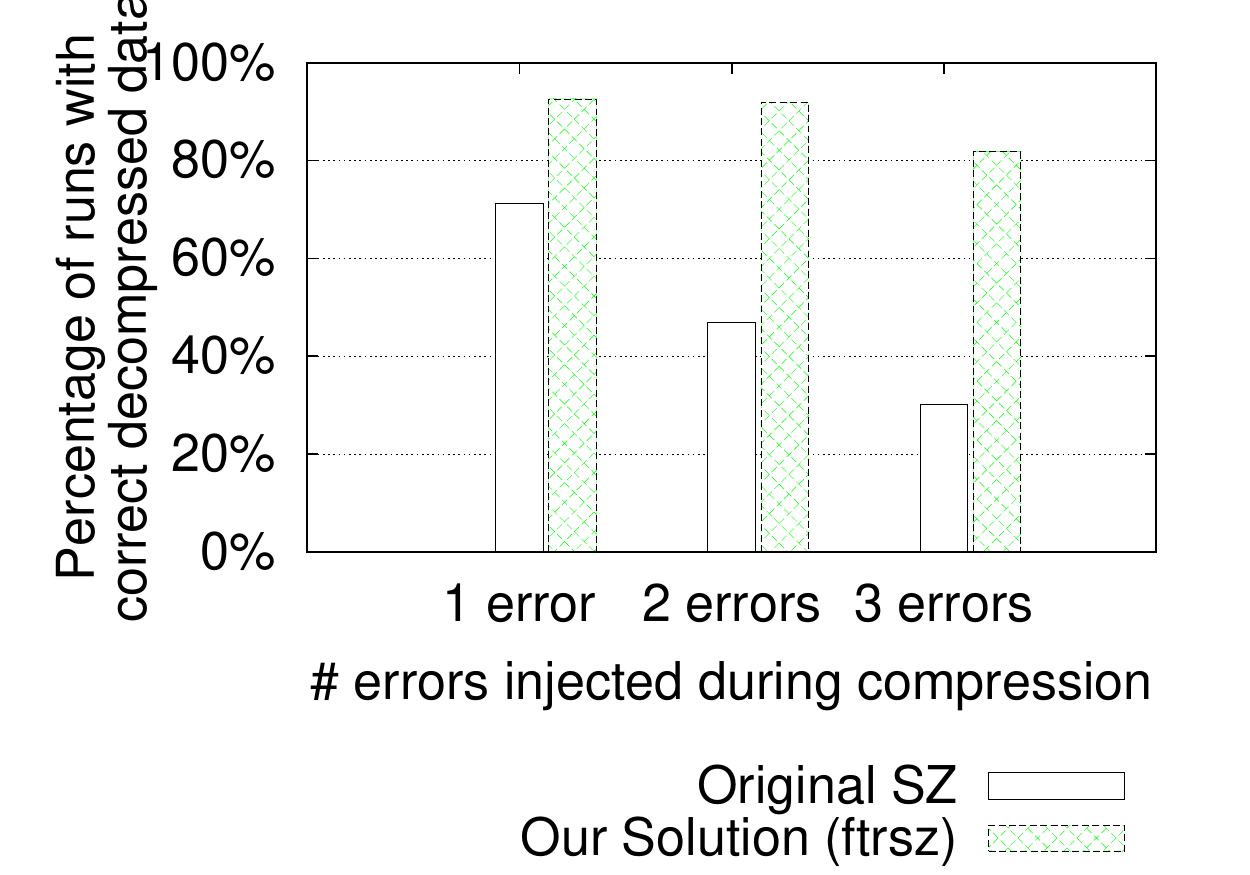}}
\hspace{-8mm}
\caption{Experimental results using evaluation mode B}
\label{fig:modeB-results}
\end{figure}

Figure \ref{fig:modeB-results} presents the experimental results of our solution (ftrsz) against the original SZ in the evaluation mode B (i.e., by injecting the errors into the whole memory during the compression). It is observed that our solution can improve the percentage of successful non-crash runs by 10\%$\sim$20\%, and improve the percentage of the runs with correct decompression results by 30\%$\sim$170\%. 
Our solution can substantially reduce the crash runs because we protect the bin arrays, which may run into core-dump segmentation faults when being injected errors, as shown in Table \ref{tab:input/type mem err}. 
In addition, as shown in Figure \ref{fig:modeB-results} (b), when injecting one and two memory errors respectively, about 92\% of running cases lead to correct decompressed data (with guaranteed error bound) under our solution, while the original SZ suffers very low percentage (71.2\% and 47\%, respectively). For our solution, the 8\% failed cases with incorrect decompression data are likely due to the error injection before the checksum execution at the beginning period, which means the checksum is calculated based on corrupted input data. Thus, it will not be able to detect future memory errors.

\subsubsection{Resilience against computation errors during compression}
\begin{figure}
    \centering
    \vspace{-6mm}
    \includegraphics[scale=.44]{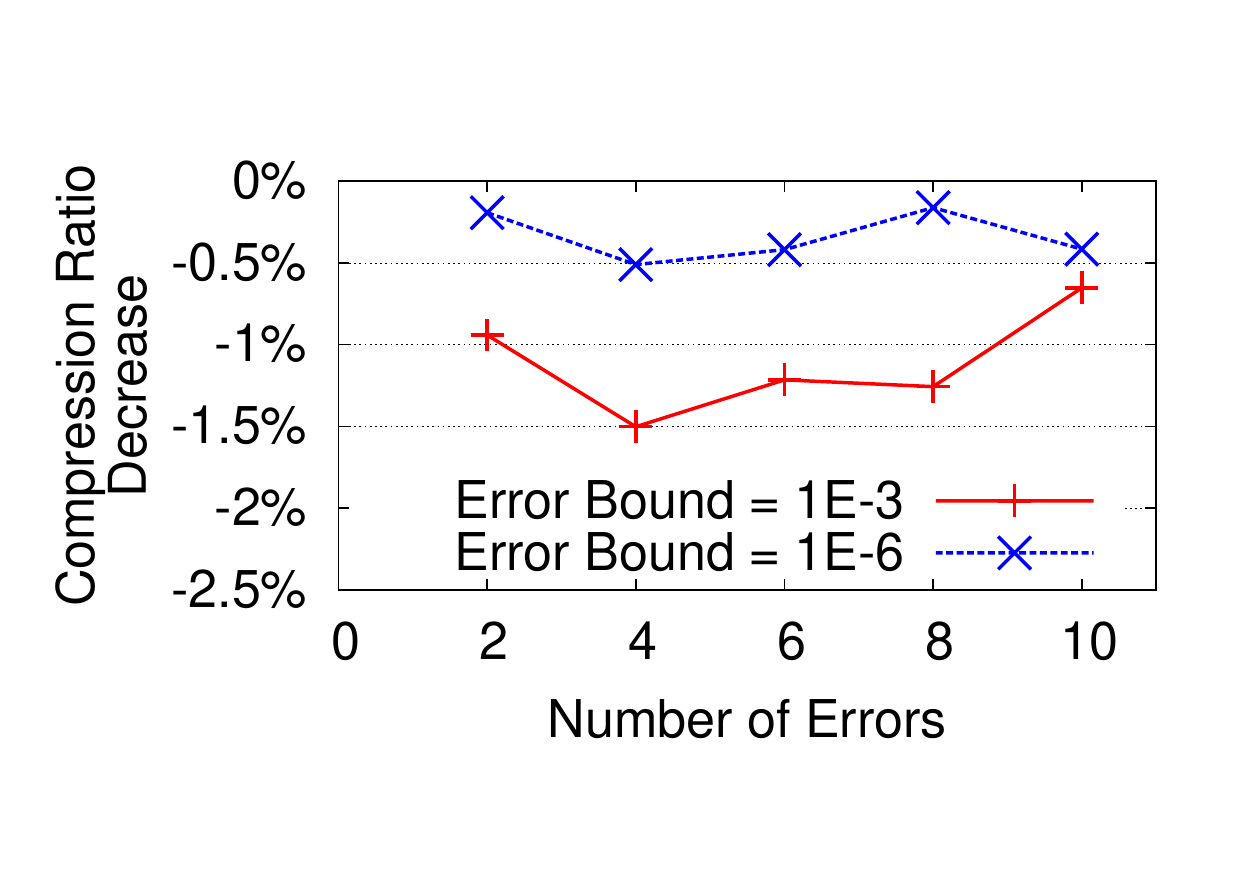}
    \vspace{-8mm}
    \caption{Compression ratio decrease with cmput. errors}
    \label{fig:compression ratio drop}
\end{figure}

As discussed in Section~\ref{sec:analysis of comp err in reg coeff}, the computations of regression coefficients, sampling and estimating compression error are error resilient though computation errors will impact the compression ratio. Figure \ref{fig:compression ratio drop} shows our experimental results about the impact to compression ratios. Computation errors are randomly injected and each experiment is repeated 50 times. The compression ratio decrease is calculated by taking the lowest compression ratio among 50 trials. As can been seen, the compression ratio decrease is within 2\% for up to 10 computation errors injected under the error bound of 1E-6 or 1E-3. The compression ratios in an error-free case are 4.8023 and 1.8112 for these two error bounds, respectively.

\subsubsection{Resilience against errors injected during decompression}

\begin{figure}[ht]
    \centering
    \subfigure[Data Dumping]{
    \includegraphics[width=4.3cm]{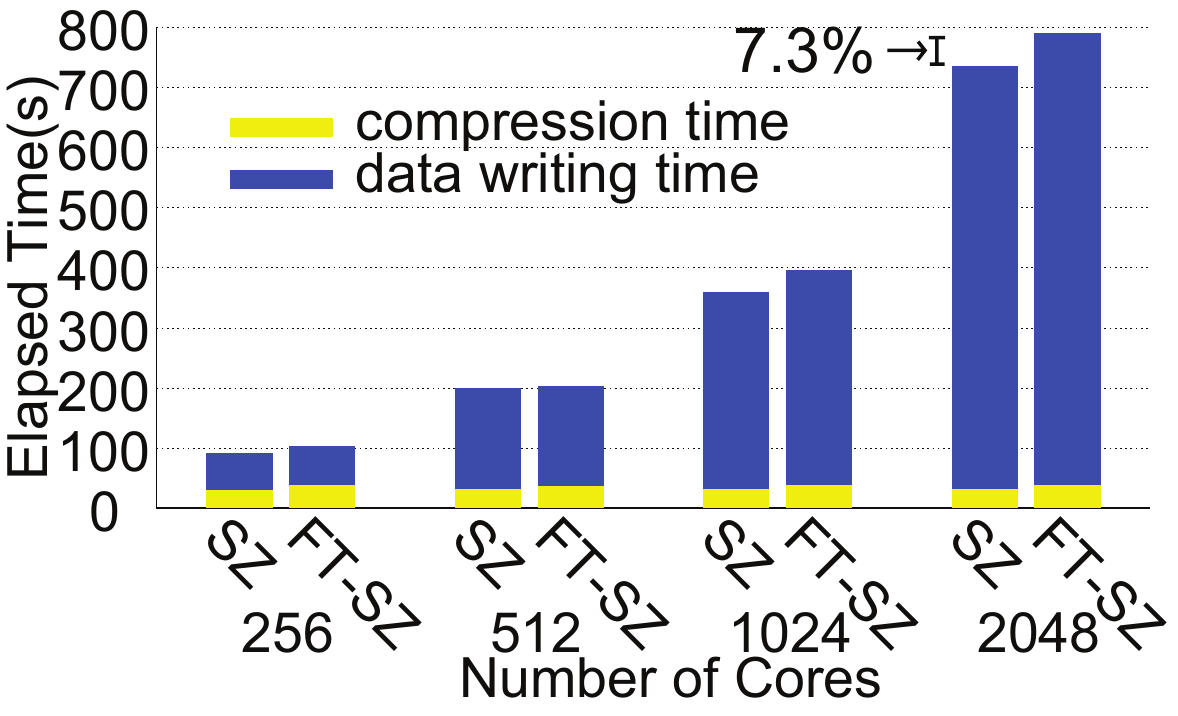}}
    \hspace{-5mm}
    \subfigure[Data Loading]{
    \includegraphics[width=4.3cm]{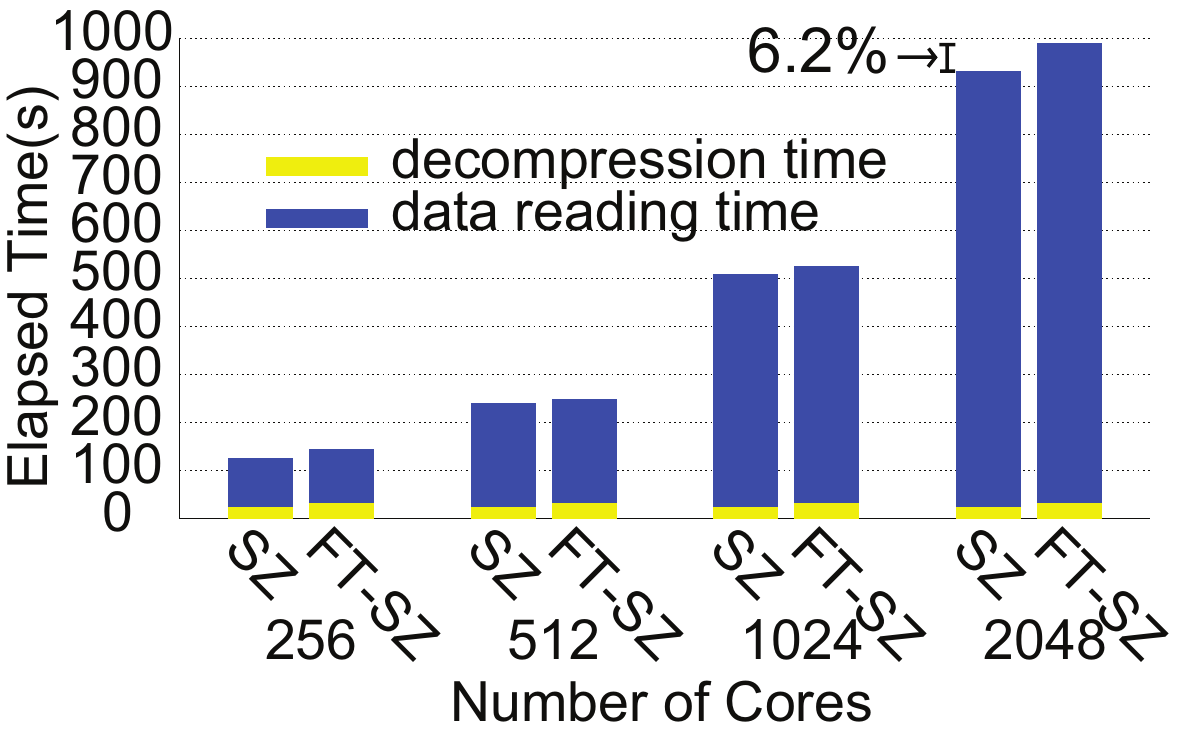}}
    
    \caption{Performance of data dumping/loading (sz vs. ftrsz)}
    \label{fig:para dumping and loading}
\end{figure}

For each run of decompression, we injected one computation error to a random block and noted all the errors can be 100\% detected by checksum and corrected by re-executing decompression for that block. Again, the extra overheads compared to fault tolerant random access SZ in error-free cases are all less than 1\% for all datasets in all error bounds.

\subsection{Parallel experimental results}

We evaluate the I/O performance with breakdown of the execution times (compression/decompression time + data writing/reading time) by processing NYX dataset under the error bound of 1E-4 in parallel on the PFS of the cluster. The experiment follows a weak-scaling style: i.e., we run the tests with different execution scales (256$\sim$2,048 cores), in which each rank kept the same data size (3GB) to process. Results are shown in Figure~\ref{fig:para dumping and loading}. As for the total data dumping time, it is observed that our error-resilient SZ incurs only 7.3\% overhead at the scale of 2,048 cores. Our error-resilient SZ has only 6.2\% overhead on the data dumping performance when using 2,048 cores to read and decompress data. The key reason for the very limited overall overhead is that the total I/O performance is dominated by compression ratio because of the I/O bottleneck of the PFS.

%% file: tex/conclusion.tex
\section{Conclusion}
In this paper, we propose a novel SDC resilient strategy for the SZ lossy compressor. We develop an independent-block based compression model for SZ to improve the robustness. We analyze each subroutine of the SZ framework elaborately and then design a series of fault tolerance strategies for the fragile code segments. We perform the evaluation by processing three well-known scientific datasets on a cluster with up to 2048 cores. Our solution can control the time overhead to about 10\%, with a degradation of compression ratio limited within about 5\%. 
When injecting one and two SDC errors respectively during the compression, our solution can have about 92\% running cases get correct decompressed data (with guaranteed error bound), which is significantly higher than that of the original SZ (71.2\% \& 47\%, respectively).
